\documentclass[useAMS,usenatbib]{mn2e}

\usepackage{graphicx}
\usepackage{amsmath}
\usepackage{nicefrac}
\usepackage{amssymb}
\usepackage{color}

\title[Discovering a 5.72 Day Period in AX J1845.0-0433]{Discovering a 5.72 Day Period in the Supergiant Fast X-ray Transient AX J1845.0-0433}
\author[M. E. Goossens et al.]{M.~E.~Goossens,$^{1}$\thanks{E-mail: M.E.Goossens@soton.ac.uk}
A.~J.~Bird,$^{1}$ S.~P.~Drave,$^{1}$ A.~Bazzano,$^{2}$ A.~B.~Hill,$^{1, 3}$
\newauthor
V.~A.~McBride,$^{4, 5}$ V.~Sguera$^{6}$ and L.~Sidoli$^{7}$ \\
$^{1}$School of Physics and Astronomy, Faculty of Physical Sciences and Engineering, University of Southampton, SO17 1BJ, UK\\
$^{2}$IAPS-INAF, Instituto di Astrofisica e Planetologia Spaziali, Via del Fosso del Cavaliere 100, I-00133 Roma, Italy\\
$^{3}$W. W. Hansen Experimental Physics Laboratory, Kavil Institute for Particle Astrophysics and Cosmology, Department of Physics and\\ SLAC National Accelerator Laboratory, Stanford University, Stanford, CA 94305, USA\\
$^{4}$Department of Astronomy, Astrophysics, Cosmology and Gravity Centre, University of Cape Town, Private Bag X3 Rondebosch, 7701,\\ South Africa\\
$^{5}$South African Astronomical Observatory, PO Box 9, Observatory, 7935, South Africa\\
$^{6}$INAF-IASF, Instituto di Astrofisica Spaziale e Fisica Cosmica, Via Gobetti 101, I-40129 Bologna, Italy\\
$^{7}$INAF-IASF, Instituto di Astrofisica Spaziale e Fisica Cosmica, Via E. Bassini 15, I-20133 Milano, Italy}

\begin{document}
\date{Accepted 2013 xx xx. Received 2013 xx xx; in original form 2013 xx xx}
\pagerange{\pageref{firstpage}--\pageref{lastpage}} \pubyear{2013}
\maketitle
\label{firstpage}
\begin{abstract}
Temporal analysis of \emph{INTEGRAL}/IBIS data has revealed a 5.7195$\pm0.0007$ day periodicity in the supergiant fast X-ray transient (SFXT) source AX J1845.0$-$0433, which we interpret as the orbital period of the system. The new-found knowledge of the orbital period is utilised to investigate the geometry of the system by means of estimating an upper limit for the size of the supergiant ($<27 R_{\sun}$) as well as the eccentricity of the orbit ($\epsilon<0.37$). 
\end{abstract}
\section{Introduction}
IGR J18450-0435 was first discovered by \emph{INTEGRAL} \citep{r22} during a survey of the Sagittarius arm tangent region carried out in spring 2003 \citep{r2}. In 2006,  \citeauthor{r3} noted that the flaring \emph{ASCA} source AX J1845.0$-$0433 and IGR J18450$-$0435 are likely to be the same object. This was later confirmed by \citet{r6} showing that the optical counterpart of \mbox{AX J1845.0$-$0433} is located 0.5\arcmin from the ISGRI position of \mbox{IGR J18450$-$0435}. 
	
AX J1845.0$-$0433 was first discovered in the Scutum arm region during an \emph{ASCA} observation in 1993 at a position of RA(2000) = $18^{\textrm{h}}$$45^{\textrm{m}}$$2^{\textrm{s}}$, Dec(2000) = -4$^\circ$33\arcmin31\arcsec. Initially, the source was in a quiescent state (0.7-10 keV flux of ~$3\times 10^{-12}$ erg $\textrm{cm}^{-2}$ $\textrm{s}^{-1}$) and then, in less than one hour, flared up so that a peak flux of $\sim10^{-9}$ erg $\textrm{cm}^{-2}$ $\textrm{s}^{-1}$ could be recorded \citep{r4}. The flat X-ray spectra obtained for the quiescent as well as the flare states both resembled those of transient X-ray binary pulsars, which led to the conclusion that AX J1845.0$-$0433 is a massive binary system consisting of a neutron star with a supergiant companion. In 1996, \citeauthor{r5} reported observations that determined the optical counterpart to \mbox{AX J1845.0$-$0433}. Optical and infrared measurements of the two brightest potential counterparts within the \emph{ASCA} error circle (1 arcmin radius) were taken and hence the counterpart was suggested to be an O9.5I supergiant star with an estimated distance of 3.6 kpc. \citet{r5}, however, pointed out that the errors in the distance could be very large mainly because of uncertainties in the reddening law. In 2008, \citeauthor{r30} reported a distance of about 7 kpc for the source, which is consistent with the new and refined value of $6.4\pm0.76$ kpc found by \citet{r33}. The high mass X-ray binary (HMXB) nature of AX J1845.0$-$0433 was fully confirmed by \citet{r6} using \emph{Swift}/XRT analysis to provide a very accurate arcsecond sized X-ray position of the source which is located only 4.7\arcsec from the supergiant star proposed by \citet{r5}. 

The fast X-ray transient behaviour, as well as its association with a supergiant star, suggests AX J1845.0$-$0433's classification as a member of the HMXB subclass of supergiant fast X-ray transients (SFXTs) \citep{r7, r20}. SFXTs are occasionally observed in X-ray quiescence ($L_x<10^{32} \textrm{erg} \textrm{s}^{-1}$) but more frequently during a low X-ray state with $L_x=10^{33}-10^{34} \textrm{erg} \textrm{s}^{-1}$. Rarely, however, fast X-ray transient activity can be observed from these sources; often lasting less than a day and being characterised by flares with short timescales of a few tens of minutes with peak luminosities of $\sim10^{36}$ erg $\textrm{s}^{-1}$. The lowest hard X-ray emission from AX J1845.0$-$0433 that was seen outside flares by \emph{INTEGRAL} (22-50 keV energy band) is $1.5\times10^{-11}$erg $\textrm{cm}^{-2}$$\textrm{s}^{-1}$ \citep{r37}. 

The most recent outburst activity from \mbox{AX J1845.0$-$0433}, detected by \emph{Swift}, was reported by \citet{r32} occurring just after two outbursts in 2010 \citep{r27}; identified by \emph{INTEGRAL}. Moreover, an \emph{XMM} observation in April 2006 detected two bright and short flares reaching $L_{x}\sim8\times10^{36}\textrm{erg} \textrm{s}^{-1}$ while the variability factor throughout the entire observation was about 50 on timescales as short as hundreds of seconds \citep{r25}. Since outbursts from SFXTs are quite rare and long periods of inactivity occur it is difficult to identify periodicities. Periodic flaring of these sources is very likely to represent the orbital period of the binary system consequently giving more indications about their geometry \citep{r1}. This in return is important for understanding the evolution of these systems and the way that mass transfer has taken place. To date coherent X-ray pulsations have not been detected from AX J1845.0$-$0433 in the range $\sim0.1-5000$ s \citep{r4}.

Here we report the temporal study of the supergiant fast X-ray transient AX J1845.0$-$0433 leading to the discovery of a 5.72 day signal we interpret as the binary orbital period.

\section{Data Set and Analysis}

The \emph{INTEGRAL} gamma-ray observatory consists of three co-aligned coded mask telescopes: the soft X-ray monitor (JEM-X)\citep{r11}, the spectrometer (SPI)\citep{r12} and the hard X-ray imager (IBIS)\citep{r13}. The data analysis was performed using the \emph{INTEGRAL} Offline Scientific Analysis (OSA) v.9 software \citep{r23, r24}. In this paper only results from IBIS/ISGRI \citep{r29} are presented, ISGRI being one of the two detector planes that comprise IBIS.

\emph{INTEGRAL} observations are divided into science windows which usually have a duration of $\sim2000$ s. We analysed IBIS/ISGRI data for all the available science windows covering the period from 2003 March 10 to 2010 September 29 (MJD range 52708--55704) in the 18-60 keV energy band. Also, 16 science windows of public Galactic Plane Scan (GPS2) data were added ranging from 2011 May 15 to 2012 February 10 in order to allow for a greater time span to be included in the analysis. The resulting light curve was then filtered so that individual science windows where AX J1845.0$-$0433 was more than 12 degrees off-axis were excluded in order to avoid effects due to incomplete off-axis response correction \citep{r10}. Additionally, science windows with an exposure time of less than \mbox{200 s} were excluded due to the intrinsically large uncertainties on these short exposures. This resulted in an IBIS data set consisting of 4610 science windows with a total exposure time of \mbox{6.65 Ms}. 

AX J1845.0-0433 also has coverage in the \emph{RXTE} All Sky Monitor (ASM) and \emph{Swift} Burst Alert Telescope (BAT) public databases. The \emph{RXTE}/ASM dwell by dwell light curve covers MJD 50090.1 through to MJD 55846.8, with a total exposure time of $\sim4.24$ Ms. After filtering for poor quality data flags, the \emph{Swift}/BAT light curve spans from MJD 53413.0 through to MJD 55802.0 with an exposure time of $\sim18.9$ Ms.

\section{Periodicity Analysis}

\begin{figure}
	\begin{center}
		\includegraphics[width=\columnwidth]{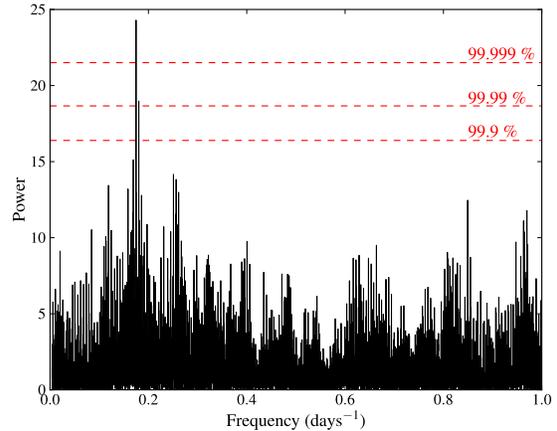}
		\caption{\label{fig:periodogram} Lomb-Scargle periodogram for the IBIS/ISGRI 18-60 keV light curve. The 99.999\% confidence level is 21.5, calculated using Monte Carlo simulations.}
	\end{center}
\end{figure}

\begin{figure}
	\begin{center}
		\includegraphics[width=\columnwidth]{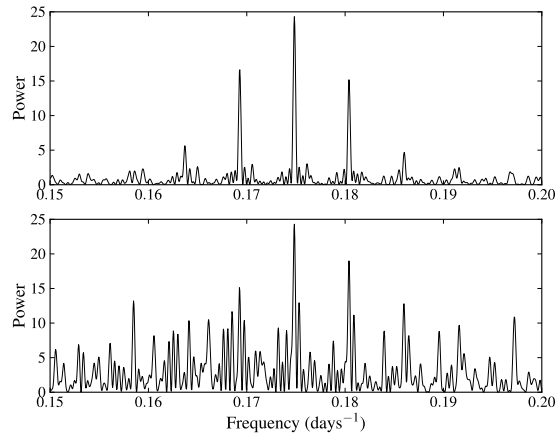}
		\caption{\label{fig:sinewave} \emph{Top}:Lomb-Scargle periodogram generated from a sine wave with a period of 5.7195 days and sampled identically to the 18-60 keV IBIS/ISGRI light curve of AX J1845.0$-$0433. \emph{Bottom}: Lomb-Scargle periodogram of AX J1845.0$-$0433 in the frequency domain, zoomed in on the peak power at a frequency of 0.174842 $\textrm{days}^{-1}$ corresponding to a period of 5.7195 days.} 
	\end{center}
\end{figure}

As a test for periodicity, the Lomb-Scargle method was used on the IBIS light curve in the 18-60 keV energy band \citep{r8, r9} using the fast implementation of \cite{r15}. A periodic signal at 0.174842 $\textrm{days}^{-1}$ could be observed, which corresponds to a periodicity of 5.7195 days. In order to determine the significance of this period Monte Carlo simulations were produced. The time stamps of the original light curve were kept the same but the flux value positions were randomised so that one can estimate the Lomb-Scargle powers expected for a random light curve with the same sampling and statistical characteristics. In order to determine the probability of randomly obtaining a peak power of $\sim24.3$ or greater, multiple runs of the analysis were needed. So as to establish a 99.999\% confidence level, 200,000 iterations were carried out and it can be seen that the probability of the observed signal being real is greater than 99.999\% (Figure \ref{fig:periodogram}). This corresponds to a significance of $4.93\sigma$, which was calculated by interpolating the corresponding significance values of the confidence levels shown in Figure \ref{fig:periodogram}. A bootstrapping analysis was completed in order to calculate the error on the period obtained in this analysis \citep{r17}. This was done by randomly selecting 70\% of the data set and performing the Lomb-Scargle analysis on it. 10,000 repetitions of this method were carried out and the distribution of the number of times a certain peak period occurred as a function of the respective period was produced. Finally, a Gaussian function was fitted and the uncertainty on the determined period is given by the standard deviation of the Gaussian curve. The best determined period is therefore 5.7195$\pm$0.0007 days.

The same periodicity testing methods were also used on the IBIS light curves in several other standard energy bands (20-40 keV, 17-30 keV, 30-60 keV, 20-100 keV). A periodic signal at 0.174842 $\textrm{days}^{-1}$  (5.7195 days) was observed in these energy bands confirming the previous result, although we point out that these do not represent truly independent test data sets. The signal could not be detected in the two higher energy bands (40-100 keV and 100-300 keV) because the source was not significantly detected at these energies.

The same periodicity was identified with a phase dispersion minimisation (PDM) analysis giving a peak power of $\sim188$ at a period of 5.719904 days using the 18-60 keV light curve. This result is consistent with the periodicity identified using the Lomb-Scargle periodogram and lies well above the corresponding 99.9\% confidence level of $\sim154$ (calculated using the same Monte Carlo simulation method as described above).

When looking at the periodogram closely it can be seen that there is some aliasing around the frequency of the peak power. Whilst being confident that the periodicity detected is the correct one due to its peak being above the relevant significance levels, this matter was investigated further in order to understand the origin of the aliasing pattern. To do so, a sine wave with the same frequency as the signal detected in the original light curve was produced and a Lomb-Scargle analysis was performed on it \citep{r21}. The corresponding periodogram as well as the one from the original data set (Figure \ref{fig:sinewave}) both exhibit the same aliasing pattern, which suggests that the detected periodicity is the real signal while additional side peaks are generated by the non-uniform sampling of the data set.

\begin{figure}
	\begin{center}
		\includegraphics[width=\columnwidth]{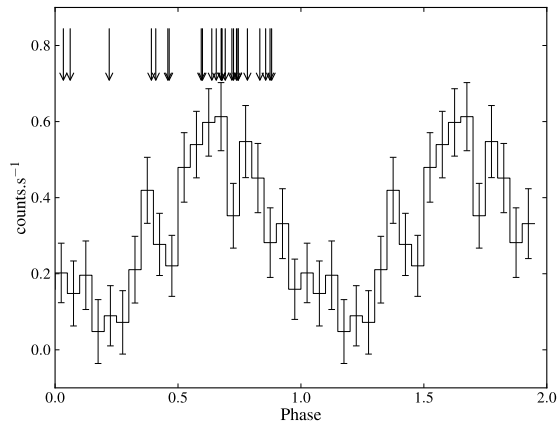}
		\caption{\label{fig:phasefold} Phase folded light curve of the IBIS/ISGRI data set for AX J1845.0$-$0433 in the 18-60 keV energy band with the ephemeris MJD 52708.43297 as phase=0 and a period of 5.7195 days. The orbital phases of the known outbursts (Table \ref{table:one}) are indicated. It can be seen that most of them are clustered around a phase of 0.6--0.7 days.}
	\end{center}
\end{figure}

Folding the 18-60 keV IBIS light curve on the 5.72-day period yields the phase folded light curve seen in Figure \ref{fig:phasefold}. It is dominated by a broad peak of an underlying flux of \mbox{0.6 counts.$\textrm{s}^{-1}$} and another two possible peaks can be observed at phases 0.35 and 0.8. The main peak occurs around phase 0.6--0.7 suggesting that this is the expected location of periastron.

Moreover, the same analysis was performed on appropriate \emph{Swift}/BAT and \emph{RXTE}/ASM data, however, no significant periodic signals could be identified. This is most likely because both instruments have lower instantaneous sensitivities than \emph{INTEGRAL}. As SFXTs have relatively low fluxes, and outburst activity is rare in these systems, it is essential to be able to detect activity over a wider flux range to allow the identification of orbital periods.

\section{Emission History}

\subsection{Outburst Identification}
\begin{figure}
	\begin{center}
		\includegraphics[width=\columnwidth]{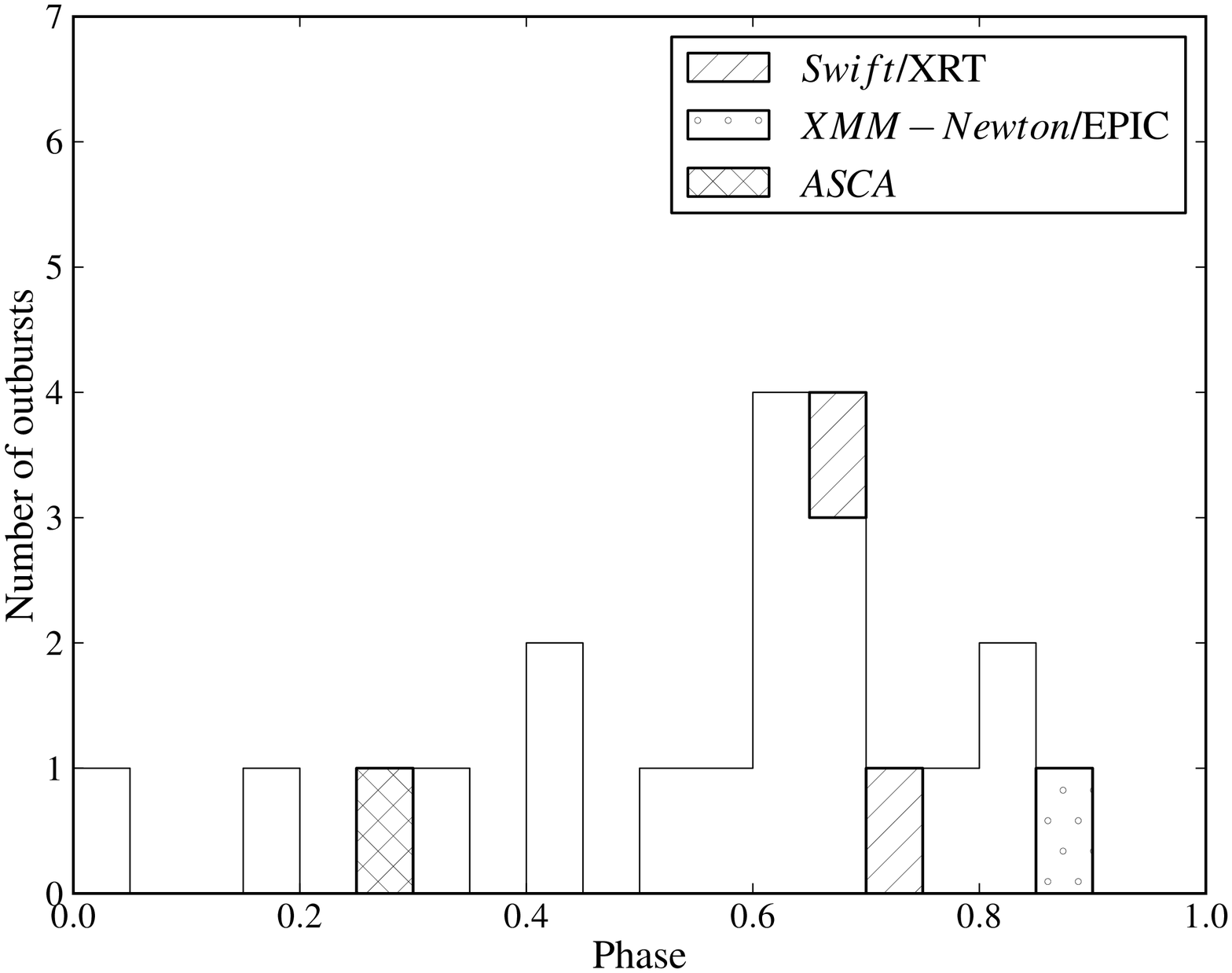}
		\caption{\label{fig:burstphase} Histogram of number of detected outbursts as a function of orbital phase. Most of the outbursts occur at  phase 0.6--0.7, which agrees with the the phase folded light curve \mbox{(Figure \ref{fig:phasefold})}.}
	\end{center}
\end{figure}

Carrying out a comprehensive search for new outbursts on the 18-60 keV light curve (a refinement to \cite{r27}) resulted in eighteen significant outbursts being identified at $5.2\sigma$ or above. Figure \ref{fig:burstphase} shows the phase location of all these outbursts together with other reported ones. It shows that most flares seem to be observed in the phase range 0.6--0.7, which coincides with the expected periastron value from the phase folded light curve (Figure \ref{fig:phasefold}). Furthermore, most of the significant outbursts seem to occur around periastron, reaching significances of around 10$\sigma$. The distribution of outbursts and the recurrence analysis in the next section both imply that there is  a level of eccentricity within the orbit as this is the likely cause of excess emission at periastron compared to apastron.

\subsection{Recurrence Analysis}
\begin{figure}
	\begin{center}
		\includegraphics[width=\columnwidth]{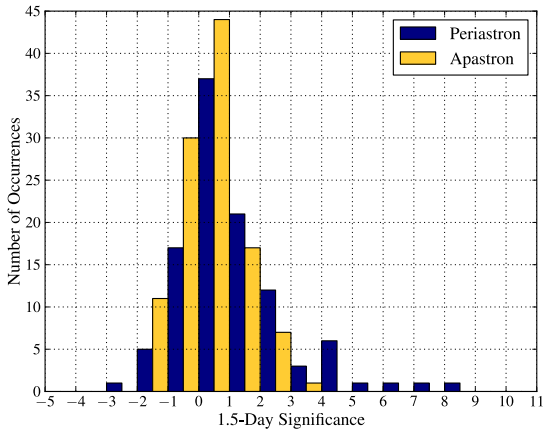}
		\caption{\label{fig:recurrence} Significance distribution for all available 1.5-day periods around periastron and apastron}
	\end{center}
\end{figure}

In order to investigate the recurrence rate for outbursts for AX J1845.0$-$0433, the IBIS light curve was analysed following the method of \cite{r1} so as to determine the 1.5-day significances (that may be either detections or non-detections) both at periastron and apastron. The significance, based simply on the weighted mean flux and error of the light curve, was calculated for $\pm0.75$ days either side of the determined periastron and apastron passages. A histogram of the distribution of significances calculated for the periastron (blue) and apastron (yellow) passages is shown in \mbox{Figure \ref{fig:recurrence}}. For the IBIS light curve, data were available for 106 possible periastron passages, and out of all the independent tests performed, 13 produced a significance over 3$\sigma$ ($\sim12\%$). When looking at the 110 apastron passages, however, only one significance was greater than $3\sigma$ ($\sim0.9\%$). This demonstrates that a markedly different significance (or flux) distribution can be observed at periastron compared to apastron. This suggests that the orbit is mildly eccentric when compared to SAX J1818.6$-$1703, for example, which displays a recurrence rate of $\sim50\%$ and is likely more eccentric \citep{r1}.

\section{Discussion}
\cite{r25} stated that to explain the \emph{XMM} source quiescent X-ray luminosity of $<2\times10^{35}$erg$\textrm{s}^{-1}$ (with a minimum luminosity as low as $1.1\times10^{34}$erg$\textrm{s}^{-1}$) a binary separation of $\textrm{a}\gtrsim5\times10^{12}$ cm (\textrm{a} $\sim3\textrm{R}_{\ast}$) is required, which is slightly higher than in classical SGXBs ($\textrm{a}\sim2\textrm{R}_{\ast}$). This suggests that the orbit of AX J1845.0$-$0433 is not very large, which agrees with the discovered orbital period. The phase-folded light curve already indicates a level of eccentricity in the system.

\subsection{Orbit Geometry and Size of the Supergiant}

Knowledge of the orbital period is very useful when trying to understand the geometry of the binary system. The 5.72-day period was used to model possible orbits and calculate a limit on the size of the supergiant.  Kepler's equation was utilised to compute the position of the neutron star around the orbit as a function of time. 
In order to find a limit on the size of the supergiant, the approximate radius of its Roche lobe was found using Eggleton's formula (\citet{r28}, Equation 1) where $r_L$ is the radius of the Roche lobe, $a$ is the semi-major axis and $q=M_1/M_2$ is the mass ratio ($M_1$ and $M_2$ being the masses of the supergiant and compact object respectively).

\begin{equation}
	\frac{r_L}{a}=\frac{0.49q^{\nicefrac{2}{3}}}{0.6q^{\nicefrac{2}{3}}+ln(1+q^{\nicefrac{1}{3}})},\qquad0<q<\infty
\end{equation}

\begin{figure}
	\begin{center}
		\includegraphics[width=\columnwidth]{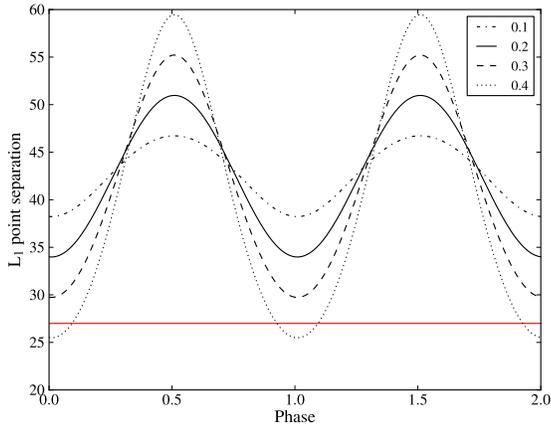}
		\caption{\label{fig:orbitfold} $\textrm{L}_{1}$ separation from supergiant against orbital phase for eccentricities ranging from 0.1-0.4. The different line styles represent the orbits in Figure \ref{fig:orbit} and the horizontal red line represents the $\sim 27R_{\sun}$ supergiant size limit.}
	\end{center}
\end{figure}

\begin{figure}
	\begin{center}
		\includegraphics[width=\columnwidth]{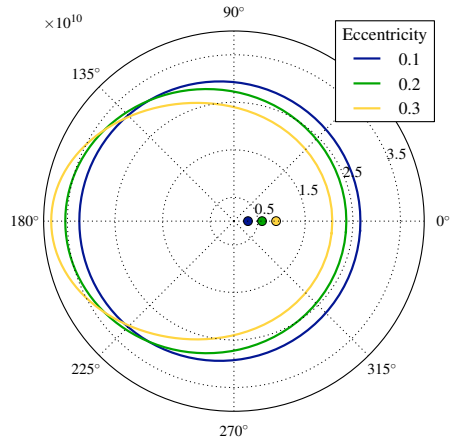}
		\caption{\label{fig:orbit} Orbit Geometry and location of supergiant for eccentricities ranging from 0.1-0.3. The locations of the respective foci where the supergiant would be are displayed by the markers towards the centre of the plot. The grid lines show the $L_{1}$ point separation from the supergiant in units of $10^{12}$ cm.}
	\end{center}
\end{figure}

Roche lobe overflow is not the driving accretion mechanism in SFXTs because it would imply a much higher mass transfer to the compact object and hence a higher X-ray luminosity than generally observed in these systems \citep{r16}. Consequently, the Roche lobe radius is also the upper limit on the supergiant radius. The Roche lobe radius was calculated using mass values of 1.4$M_{\sun}$ assuming the compact object is a neutron star and $\sim 30M_{\sun}$ for the supergiant, which is an interpolated, approximate value based upon the typical properties of supergiant stars listed in \cite{r26}. It also limits the eccentricity of the orbit because if the Lagrangian point $\textrm{L}_{1}$ recedes inside the surface of the supergiant, Roche lobe overflow would occur. Figure \ref{fig:orbitfold} shows the $L_{1}$ separation from the supergiant with respect to phase for eccentricities ranging from 0.1 to 0.4. More detailed calculations in a range of $\epsilon\sim$0.3--0.4 results in the limit being calculated as $\epsilon<0.37$. At eccentricities $\geq0.37$ the $L_{1}$ point would be so close to the supergiant that it would touch its surface inducing Roche lobe overflow. At lower eccentricities, even when the $L_{1}$ point is closest to the supergiant in that particular orbit, it is still far enough away so that no overflow can occur. In Figure \ref{fig:orbit}, three possible orbits can be seen including the locations of the foci where the supergiant would be present. The Roche lobe radius and consequently the upper limit on the size of the supergiant was calculated to be $\sim27R_{\sun}$. Observational data suggests that an O9.5I supergiant has a size of $\sim23R_{\sun}$ \citep{r26}, which is well within the Roche lobe radius.

\begin{table*}
\begin{minipage}{150mm}
\centering
\caption{Summary of \emph{INTEGRAL}, \emph{Swift}, \emph{ASCA} and XMM observations of the most significant outbursts of AX J1845.0-0433 \citep{r6, r31, r32, r16, r27} where distances of 3.6 kpc as well as 6.4 kpc were used to calculate the luminosity values. The numbers in bold font are the two refined outbursts reported in \citet{r27} and numbers 12--25 are the remaining outbursts from the outburst identification analysis (section 4.1). The flux values where calculated using the mean fluxes of each individual outburst.}
\label{table:one}
\resizebox{\columnwidth}{!}{
\begin{tabular}{@{}| c | c | c | c | c | c | c | c | c | c | }
\hline
\hline
No.&Observation&Date&Orbital&Energy Band&Flux&Luminosity&Photon Index&N$_{\textrm{H}}$&Sigificance\\
&&& Phase&(keV)&(erg $\textrm{cm}^{-2} \textrm{s}^{-1}$)&(erg $\textrm{s}^{-1}$)&(Power Law)&($\textrm{cm}^{-2}$)&($\sigma$)\\
\hline
1&\emph{ASCA}&18 Oct. 1993&0.25&0.7-10&$1.0\times10^{-9}$&$(1.5-4.9)\times10^{36}$&1.0$^{+0.07}_{-0.07}$&$3.6\pm0.3\times10^{22}$&\\ [2pt]
2&\emph{INTEGRAL}/ISGRI&28 Apr. 2005&0.30&20-40&$4.5\times10^{-10}Ê$&$7.0\times10^{35}-2.2\times10^{36}$&2.5$^{+0.6}_{-0.5}$&&$\sim6$\\[2pt] 
3&\emph{Swift}/XRT&11 Nov. 2005&0.74&0.2-10&$2.3\times10^{-10}$&$3.6\times10^{35}-1.1\times10^{36}$&0.75$^{+0.1}_{-0.1}$&$1.6\pm0.18\times10^{22}$&\\[2pt]
4&\emph{Swift}/XRT&05 Mar. 2006&0.68&0.2-10&$1.1\times10^{-10}$&$(2.0-5.4)\times10^{35}$&0.85$^{+0.3}_{-0.3}$&$2.3\pm0.7\times10^{22}$&\\[2pt]
5&\emph{XMM-Newton}/EPIC&03 Apr. 2006&0.86&0.4-10&$1.2\times10^{-9}$&$(1.9-5.9)\times10^{36}$&$0.8\pm0.1$&$2.6\pm0.2\times10^{22}$&\\[2pt]
6&\emph{INTEGRAL}/ISGRI&20 Apr. 2006&0.72&20-40&$6.0\times10^{-10}$&$9.3\times10^{35}-2.9\times10^{36}$&2.9$^{+0.9}_{-0.7}$&&$\sim6$\\[2pt]
7&\emph{INTEGRAL}/ISGRI&03 Sept. 2006&0.66&18-60&$2.5\times10^{-10}$&$3.9\times10^{35}-1.2\times10^{36}$&2.62$^{+0.4}_{-0.4}$&&\\[2pt]
8&\emph{Swift}/BAT&28 June 2009&0.41&15-150&$1.5\times10^{-9}$&$(2.4-7.4)\times10^{36}$&2.41$^{+0.34}_{-0.34}$&&\\[2pt]
&\emph{Swift}/XRT&28 June 2009&0.41&0.3-10&$3.0\times10^{-10}$&$4.7\times10^{35}-1.5\times10^{36}$&&&\\[2pt]
\bf{9}&\emph{INTEGRAL}/ISGRI&14 Mar. 2010&0.83&18-60&$2.4\times10^{-10}$&$3.7\times10^{35}-1.2\times10^{36}$&2.53$^{+0.3}_{-0.3}$&&9.55\\[2pt]
\bf{10}&\emph{INTEGRAL}/ISGRI&20 Mar. 2010&0.88&18-60&$4.0\times10^{-10}$&$6.2\times10^{35}-2.0\times10^{36}$&2.84$^{+0.3}_{-0.3}$&&11.10\\[2pt]
11&\emph{Swift}/BAT&05 May 2012&0.59&15-150&$3.0\times10^{-9}$&$4.7\times10^{36}-1.5\times10^{37}$&2.47$^{+0.53}_{-0.53}$&&\\[2pt]
&\emph{Swift}/XRT&05 May 2012&0.59&2-10&$7.0\times10^{-10}$&$(1.1-3.4)\times10^{36}$&1.5$^{+0.1}_{-0.1}$&1.8$^{+0.3}_{-0.1}\times10^{22}$&\\[2pt]
\hline
12&\emph{INTEGRAL}/ISGRI&06 Apr. 2003&0.68&18-60&$1.6\times10^{-10}$&$(2.4-7.8)\times10^{35}$&&&6.11\\[2pt]
13&\emph{INTEGRAL}/ISGRI&23 Apr. 2003&0.74&18-60&$1.7\times10^{-10}$&$(2.6-8.3)\times10^{35}$&&&5.42\\[2pt]
14&\emph{INTEGRAL}/ISGRI&06 May 2003&0.03&18-60&$2.7\times10^{-10}$&$4.2\times10^{35}-1.3\times10^{36}$&&&5.25\\[2pt]
15&\emph{INTEGRAL}/ISGRI&30 Sept. 2003&0.74&18-60&$2.8\times10^{-10}$&$4.3\times10^{35}-1.4\times10^{36}$&&&6.04\\[2pt]
16&\emph{INTEGRAL}/ISGRI&10 Oct. 2004&0.46&18-60&$8.4\times10^{-11}$&$(1.3-4.1)\times10^{35}$&&&5.78\\[2pt]
17&\emph{INTEGRAL}/ISGRI&29 Oct. 2004&0.69&18-60&$2.6\times10^{-10}$&$4.0\times10^{35}-1.3\times10^{36}$&&&10.22\\[2pt]
18&\emph{INTEGRAL}/ISGRI&23 Apr. 2005&0.60&18-60&$9.1\times10^{-11}$&$(1.4-4.5)\times10^{35}$&&&8.85\\[2pt]
19&\emph{INTEGRAL}/ISGRI&13 May 2005&0.22&18-60&$2.1\times10^{-10}$&$3.3\times10^{35}-1.0\times10^{36}$&&&5.73\\[2pt]
20&\emph{INTEGRAL}/ISGRI&25 Apr. 2006&0.68&18-60&$9.5\times10^{-11}$&$(1.5-4.7)\times10^{35}$&&&5.23\\[2pt]
21&\emph{INTEGRAL}/ISGRI&24 Aug. 2007&0.64&18-60&$4.6\times10^{-10}$&$7.1\times10^{35}-2.3\times10^{36}$&&&5.66\\[2pt]
22&\emph{INTEGRAL}/ISGRI&10 Oct. 2007&0.87&18-60&$1.1\times10^{-10}$&$(1.7-5.4)\times10^{35}$&&&6.40\\[2pt]
23&\emph{INTEGRAL}/ISGRI&19 Oct. 2007&0.39&18-60&$2.1\times10^{-10}$&$3.3\times10^{35}-1.0\times10^{36}$&&&5.29\\[2pt]
24&\emph{INTEGRAL}/ISGRI&02 Mar. 2008&0.06&18-60&$1.5\times10^{-10}$&$(2.3-7.4)\times10^{35}$&&&5.21\\[2pt]
25&\emph{INTEGRAL}/ISGRI&23 Sept. 2009&0.73&18-60&$3.6\times10^{-10}$&$5.6\times10^{35}-1.8\times10^{36}$&&&5.36\\[2pt]

\end{tabular}}
\end{minipage}
\end{table*}

\subsection{The Nature of AX J1845.0$-$0433}

AX J1845.0$-$0433 is now firmly established as a member of the SFXT class, and the determination of the orbital period enables the emissions seen over 20 years to be placed in a more robust orbital context. 

In the case of AX J1845.0$-$0433, the hard X-ray emission appears to be dominated by low-luminosity accretion that overall produces a marked orbital variation. Hence the discovery of the orbital period has required the accumulation of considerable amounts of observational data. With the possible exception of the `discovery outburst' in 1993, there is little evidence of the larger outbursts seen in some systems such as SAX J1818.6$-$1703 while the detected outbursts show the same orbit profile as the folded light curve from all data. AX J1845.0$-$0433 has a dynamic range of about 40 above 20 keV and about 400 in the 0.2-10 keV energy band, which is significantly lower than that of classical archetypal SFXTs. This suggests that AX J1845.0$-$0433 is likely to be an intermediate SFXT similar to IGR J16465$-$4507 \citep{r34}.

In the context of the clumpy wind model, we can view AX J1845.0$-$0433 as an SFXT, with a mildly eccentric orbit bringing the neutron star within the denser, clumpier wind of its supergiant companion every 5.72 days, with a higher probability of flaring activity resulting. However, more recent reviews suggest that outbursts from systems with shorter orbital periods like this one might be better explained by an alternate model due to their very different outburst rates compared to systems with longer orbital periods \citep{r36}. Understanding a potentially new mechanism that reduces the accretion rate onto the compact object might be required \citep{r35}. \mbox{AX J1845.0$-$0433} is distant enough that hard X-ray instrumentation such as \mbox{\emph{INTEGRAL}}/IBIS only sees the `tip of the iceberg' in terms of this flaring, and a much larger population of sub-threshold flares are likely to contribute to its orbital flux profile.

\section*{Acknowledgements}
M. E. Goossens is supported by a Mayflower Scholarship from the University of Southampton. S. P. Drave acknowledges support from the UK Science and Technology Facilities Council, STFC. V. A. McBride acknowledges financial support from the NRF, South Africa, and the World Universities Network. A.~B. Hill acknowledges that this research was supported by a Marie Curie International Outgoing Fellowship within the 7th European Community Framework Programme (FP7/2007--2013) under grant agreement no. 275861

\label{lastpage}
\end{document}